\documentstyle[preprint,tighten,aps,epsfig]{revtex}
\def\csw{c_{\rm sw}}
\def\psibar{\overline{\psi}}
\begin{document}
\preprint{\parbox{5cm}{DESY 01-084\\
Edinburgh-2001/05\\ HU-EP-01/24\\JLAB-THY-01-017}}
\title{Negative-Parity Baryon Masses using an ${\cal O}(a)$-Improved
Fermion Action}
\author{
M.~G\"{o}ckeler$^1$, R.~Horsley$^{2,3}$, 
D.~Pleiter$^2$, P.E.L.~Rakow$^1$, \\G.~Schierholz$^{2,4}$
(QCDSF Collaboration)\\[1.0ex]
C.M.~Maynard$^5$ (UKQCD Collaboration)\\[1.0ex]
D.G.~Richards$^{6,7}$ (LHPC Collaboration)}
\address{
$^1$Institut f\"ur Theoretische Physik, Universit\"at Regensburg,
                    D-93040 Regensburg,
                    Germany\\
$^2$Deutsches Elektronen-Synchrotron DESY,
             John von Neumann Institute for Computing NIC/Deutsches
	     Elektronen-Synchrotron DESY, \\
                    D-15738 Zeuthen, Germany\\
$^3$Institut f\"ur Physik, Humboldt-Universit\"at zu Berlin,
                    D-10115 Berlin, Germany \\
$^4$Deutsches Elektronen-Synchrotron DESY,
                    D-22603 Hamburg, Germany\\
$^5$Department of Physics \& Astronomy, University of Edinburgh,
Edinburgh EH9~3JZ, Scotland, UK\\
$^6$Jefferson Laboratory, MS 12H2, 12000 Jefferson Avenue, 
Newport News, VA 23606, USA\\
$^7$Department of Physics, Old Dominion University, Norfolk, 
        VA 23529, USA}
\maketitle
\begin{abstract}
We present a calculation of the mass of the lowest-lying
negative-parity $J=1/2^{-}$ state in quenched QCD.  Results are
obtained using a non-perturbatively ${\cal O}(a)$-improved clover
fermion action, and a
splitting is found between the mass of the nucleon, and its parity
partner.  The calculation is performed on two lattice volumes and at
three lattice spacings, enabling a study of both finite-volume and
finite lattice-spacing uncertainties.  A comparison is made with
results obtained using the unimproved Wilson fermion action.
\end{abstract}
\narrowtext
\section{Introduction}
The study of the excited nucleon spectrum can provide important clues
to the dynamics of QCD and the nature of the interactions between
its fundamental partons. The observed $N^*$ spectrum raises many
important questions, such as the nature of the Roper resonance, and
whether the $\Lambda(1405)$ is a true three-quark state or a molecular
state.  For these reasons, the study of the spectrum is an important element of
the Jefferson Laboratory experimental programme.  The phenomenological
interest in the excited nucleon spectrum has been complemented by a
flurry of activity in the lattice community.  In particular, two
calculations of the mass of the parity partner of the nucleon have
appeared; the first employed the highly-improved $D_{234}$ fermion
action\cite{lee98,lee00}, whilst the second employed domain-wall
fermions\cite{sasaki99,sasaki01}. Both calculations exhibited a clear
splitting between the masses of the $N^{1/2+}$ and $N^{1/2-}$ states,
and the importance of chiral symmetry breaking in obtaining a non-zero mass
splitting has been stressed~\cite{sasaki99}.  

In this paper, we present a calculation of the lowest lying negative
parity nucleon using an ${\cal O}(a)$-improved Sheikholeslami-Wohlert
(SW), or clover, fermion action; preliminary results were presented in
ref.~\cite{lat00}.  By choosing the coefficient of the improvement term
appropriately, all ${\cal O}(a)$ discretisation uncertainties are
removed, ensuring that the continuum limit is approached with a rate
proportional to $a^2$.

The calculation of the excited nucleon spectrum places particularly
heavy demands on lattice spectroscopy.  The excited nucleon states are
expected to be large; the size of a state is expected to double with
each increase in orbital angular momentum.  Thus a lattice study of
the excited nucleon spectrum requires large lattice volumes, with
correspondingly large computational requirements.  Furthermore, the
states are relatively massive, requiring a fine lattice spacing, at
least in the temporal direction.  These requirements could be
satisfied with much greater economy using the clover fermion action
than using the domain-wall or overlap formulation.  Thus it is
important to establish that the negative parity states are indeed
accessible to calculations using the clover action.  Finally, by
comparing the masses obtained using the clover action with a
calculation, at a single quark mass, using the Wilson fermion action,
we also gain insight into the nature of the interaction responsible for the
splitting in the parity doublet.

The rest of the paper is laid out as follows.  In the next section, we
introduce the hadronic operators and correlators measured in the
calculation, describe the fermion action, and provide the simulation
parameters.  Section~\ref{sec:results} contains our results for the
masses of the lowest lying positive- and negative-parity nucleon
states using the SW fermion action.  Detailed discussion,
including a comparison with the Wilson fermion case, and our
conclusions, are presented in Section~\ref{sec:concl}.

\section{Calculational Details}\label{sec_operators}
\subsection{Baryon Operators}
For a particle at rest, there are three local interpolating operators
that that have an overlap with the nucleon:
\begin{eqnarray}
N_1^{1/2+} & = & \epsilon_{ijk} (u_i^T C \gamma_5 d_j) u_k\label{eq:N1},\\
N_2^{1/2+} & = & \epsilon_{ijk} (u_i^T C d_j) \gamma_5 u_k\label{eq:N2},\\
N_3^{1/2+} & = & \epsilon_{ijk} (u_i^T C \gamma_4 \gamma_5 d_j) u_k.
\label{eq:N3}
\end{eqnarray}
The ``diquark'' part of both $N_1$ and $N_3$ couples upper spinor
components, while that in $N_2$ involves the lower components and thus
vanishes in the non-relativistic limit\cite{lee98}.  For some of the
lattices in our calculation, the positive-parity nucleon mass is
obtained using the non-relativistic quark operators~\cite{qcdsf95}, defined by
\begin{equation}
\psi \rightarrow \psi^{\rm NR} = \frac{1}{2} ( 1 + \gamma_4) \psi,\,
\overline{\psi}^{\rm NR} = \overline{\psi} \frac{1}{2} (1 + \gamma_4).
\end{equation}
In practice, lattice calculations confirm the naive expectation that
the operators $N_1$ and $N_3$ have a much greater overlap with the
nucleon ground state than $N_2$, and therefore we do not use this
operator in the fits.

The correlators constructed from these operators receive contributions
from both the positive-parity nucleon, and from its (heavier) negative
parity partner.  However, we can achieve some delineation of the
states into forward-propagating positive-parity states and
backward-propagating negative-parity states, or the converse, through
the use of the parity projection operator $(1 \pm \gamma_4)$.  On a
lattice periodic or anti-periodic in time, the resulting correlators
may be written:
\begin{eqnarray}
\lefteqn{
C_{N_i^{+/-}}(t) = \sum_{\vec{x}} \left( (1\pm\gamma_4)_{\alpha\beta}
\langle N_{i,\alpha}(\vec{x},t) \overline{N}_{i,\beta}(0) \rangle +
\right.} \nonumber\\
& & \left. (1 \mp \gamma_4)_{\alpha\beta} \langle N_{i,\alpha}(\vec{x}, N_t -
t) \overline{N}_{i,\beta}(0)\rangle \right),\label{eq:corrs}
\end{eqnarray}
where $N_t$ is the temporal extent of the lattice.
At large distances, when $t \gg 1$ and $N_t -t \gg 1$, the correlators
behave as
\begin{eqnarray}
C_{N_i^+}(t) & \rightarrow & A_i^+ e^{-m_i^+ t} + A_i^- e^{-m_i^-(N_t
- t)} \label{eq:corr_pos}\\
C_{N_i^-}(t) & \rightarrow & A_i^- e^{-m_i^- t} + A_i^+ e^{-m_i^+(N_t -
t)}\label{eq:corr_neg}
\end{eqnarray}
where $m_i^+$ and $m_i^-$ are the lightest positive- and
negative-parity masses respectively in channel $i$.  

\subsection{Fermion Action}
To leading order in $a$ the Symanzik improvement programme amounts to adding
the well-known Sheikholeslami-Wohlert term to the fermionic Wilson action
\cite{SW85}
\begin{equation}
  \delta S = -\csw\frac{i\kappa}{2}\sum_{x,\mu,\nu}\psibar(x)
  \sigma_{\mu\nu}F_{\mu\nu}(x) \psi(x).\label{eq:sw_action}
\end{equation}
Provided that $\csw$ is chosen appropriately, spectral quantities such
as hadron masses approach the continuum limit with a rate proportional
to $a^2$. Non-perturbative determinations of $\csw$ have been made in
the quenched approximation to QCD in refs.~\cite{AlphaIII} and
\cite{SCRI_imp}.

The Sheikholeslami-Wohlert term, eqn.~(\ref{eq:sw_action}), is of
magnetic moment form, and it is well known that the use of the SW
action results in hyperfine splittings that are closer to their
experimental values than those obtained using the standard Wilson
fermion action, see, for example, ref.~\cite{light93}.  The SW term
also removes the leading chiral-symmetry-breaking effects at finite
$a$.  In view of these considerations, we compare the splitting
between the positive- and negative-parity states obtained using the
Wilson fermion action at a single light-quark mass with the results
obtained using the SW fermion action.

\subsection{Simulation Details}
The calculation is performed in the quenched approximation to QCD
using lattices generated by the UKQCD and QCDSF collaborations;
calculations of the light hadron spectrum using these lattices have
appeared in ref.~\cite{UKQCD} and \cite{QCDSF,pleiter00} respectively.
Propagators on the UKQCD lattices were computed from both local and
fuzzed sources to both local and fuzzed sinks; the fuzzing procedure
is described in ref.~\cite{fuzzing}.  The parameters used in the
calculation are listed in Table~\ref{tab:params}.
Propagators on the QCDSF lattices were computed using Jacobi smearing
at both source and sink, described in ref.~\cite{SFsmear}.

The errors on the fitted masses are computed using a bootstrap
procedure.  In the case of the UKQCD data, the same number of
configurations are used for each of the quark masses for a given
$\beta \equiv 6/g^2$ and lattice volume.  However, in the case of the
QCDSF configurations, different numbers of configurations are used at
different quark masses even at the same $\beta$ and volume.  This
precludes the use of correlated fits in the chiral extrapolations, and
thus a simple uncorrelated $\chi^2$ fit is performed, with the
uncertainties computed from the variation in the $\chi^2$.

\section{Results}\label{sec:results}
\begin{figure}[tp]
\begin{center}
\epsfig{file=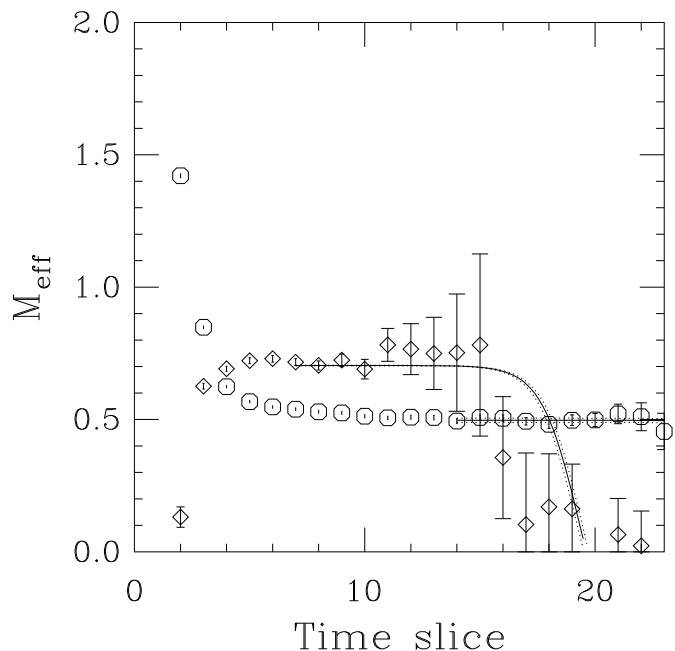,width=250pt}
\end{center}
\caption{The effective masses for the $N^{1/2+}$ channel (circles) and
the $N^{1/2-}$ channel (diamonds) at $\beta = 6.2$
with $\kappa= 0.1351$.  The lines are from a simultaneous fit to both
parities.}
\label{fig:ukqcd_62_eff_mass}
\end{figure}

\begin{figure}[tp]
\begin{center}
\epsfig{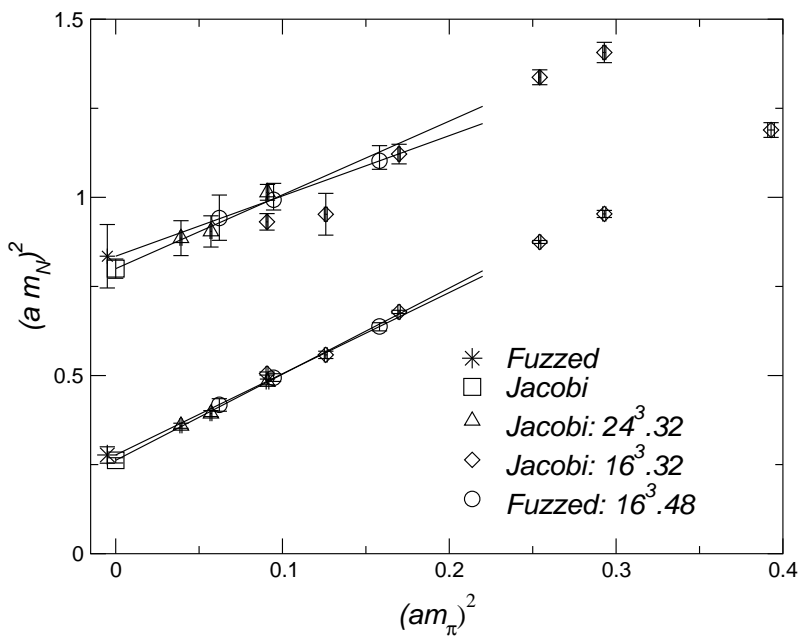}
\end{center}
\caption{The masses in lattice units of the lowest-lying positive- and
negative- parity nucleons on the lattices at $\beta
=6.0$.  The curves are from independent fits to the Jacobi-smeared and
fuzzed data for $m_X^2$ using
eqn.~(\ref{eq:chiral_extrap}), as described in the text.}
\label{fig:beta60_all}
\end{figure}

\begin{figure}[tp]
\begin{center}
\epsfig{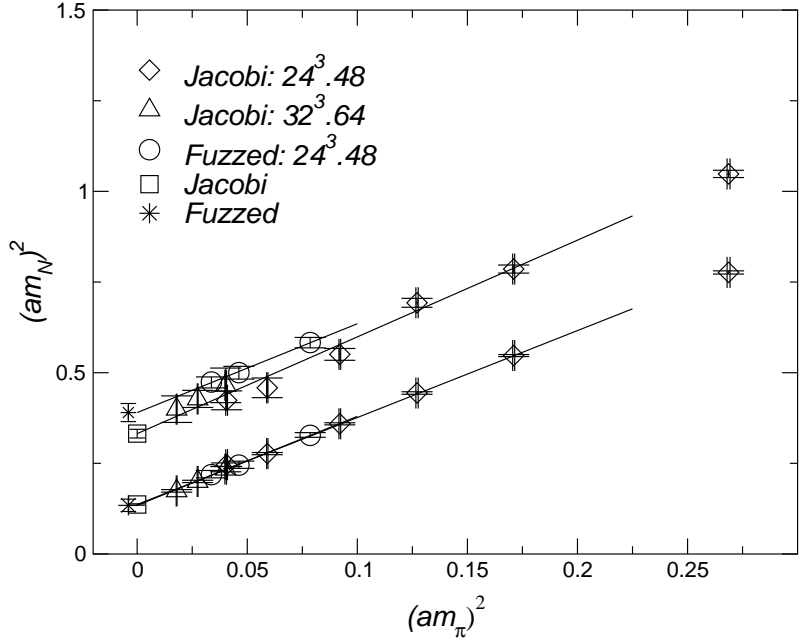}
\end{center}
\caption{The masses in lattice units of the lowest-lying positive- and
negative-parity nucleons at $\beta = 6.2$.  The curves are from
independent fits to the Jacobi-smeared and fuzzed data for $m_X^2$
using eqn.~(\ref{eq:chiral_extrap}).}
\label{fig:beta62_all}
\end{figure}

\begin{figure}[tp]
\begin{center}
\epsfig{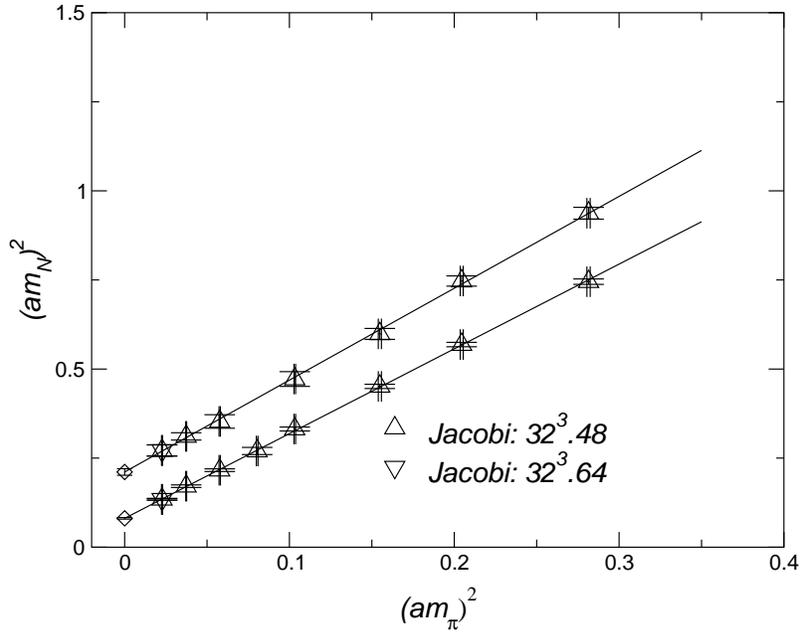}
\end{center}
\caption{The masses in lattice units of the lowest-lying positive- and
negative-parity nucleons at $\beta
=6.4$. The curves are from fits to $m_X^2$ using
eqn.~(\ref{eq:chiral_extrap}).}
\label{fig:beta64_all}
\end{figure}

The masses of the lowest-lying $N^{1/2+}$ and $N^{1/2-}$ states are
obtained from a simultaneous, four-parameter fit to the positive- and
negative-parity correlators of eqn.~(\ref{eq:corrs}), constructed
using fuzzed sources and local sinks (UKQCD) or using smeared sources
and smeared sinks (QCDSF), using the fit functions of
eqns.~(\ref{eq:corr_pos}) and (\ref{eq:corr_neg}). We see a clear
signal for the mass of the negative-parity states, and the quality of
a fit is illustrated in Figure~\ref{fig:ukqcd_62_eff_mass}; the
contamination of the negative-parity correlator from the lighter,
backward-propagating, positive-parity state is clear both in the fits
and in the data.  The masses of the lightest particle of positive and
negative parity as a function of $m_{\pi}^2$ on each ensemble are
shown in Figures \ref{fig:beta60_all}-\ref{fig:beta64_all}.

For the chiral extrapolations of the hadron masses, we adopt the
ansatz
\begin{equation}
(a m_X)^2 = (a M_X)^2 + b_2 (a m_{\pi})^2\label{eq:chiral_extrap}
\end{equation}
where we use upper-case letters to denote masses obtained in the
chiral limit, and $X$ is either $N^{1/2+}$ or $N^{1/2-}$.  We include
data at different volumes, but at the same $\beta$, in the chiral
extrapolations, but treat the fuzzed and Jacobi-smeared data
independently; we include only those masses which are less than of
order one in lattice units.  The parameters of the fit for the
positive- and negative-parity states are given in
Table~\ref{tab:chiral_extrap}, and the chiral extrapolation, together
with the extrapolated masses, shown in
Figures~\ref{fig:beta60_all}-\ref{fig:beta64_all}.  Note that the fit
to $m_X^2$, rather than $m_X$, gives a sensible behaviour in the
heavy-quark limit, whilst being formally the same at light
pseudoscalar masses; a plot of $m_X$ vs.\ $m_{\pi}^2$ at $\beta = 6.4$
exhibits clear curvature.

From quenched chiral perturbation theory one expects leading
non-analytic terms which are linear in $m_{\pi}$~\cite{labrenz96}. The
coefficient of the leading term is predicted to be negative. We
therefore also attempted to fit our data at $\beta = 6.4$ to the form
\begin{equation}
(am_X)^2 = (a M_X)^2 + b_1 (a m_{\pi}) + b_2 (a m_{\pi})^2.
\label{eq:non_analytic}
\end{equation}
Fitting the $N^{1/2+}$ state we found a positive value of $b_1$,
though with a very large error that would still accommodate a negative
value. Thus it is unclear whether eqn.~\ref{eq:non_analytic} provides
a reliable form with which to extrapolate data obtained with quark
masses around that of the strange quark.  Furthermore, the chirally
extrapolated values thus obtained differ from those obtained using
eqn.~\ref{eq:chiral_extrap} by only around 5\%. We therefore
quote chirally extrapolated masses from the fit to
eqn.~\ref{eq:chiral_extrap} in the remainder of this letter.  There
has been considerable study of the contribution of non-analytic terms
arising from pion-induced baryon self energies~\cite{leinweber99}.  We
do not investigate the effect of these contributions in this paper.

The ideal procedure for obtaining the chirally extrapolated masses
would be to apply the forms of eqn.~\ref{eq:chiral_extrap} and
\ref{eq:non_analytic} to the infinite-volume limit of the baryon
masses obtained at each pseudoscalar mass.  Unfortunately, our data do not
allow this procedure, but we try to gain some insight into the
possible finite-volume uncertainties by comparing the masses otained
at a value of the pseudoscalar mass for which we have data on both
smaller and larger volumes.  Specifically, we compare the masses
obtained from the Jacobi-smeared data at $\beta = 6.0$ and $\beta =
6.2$ with a pseudoscalar mass given by
\begin{equation}
a m_{\pi} = 4.8/(L/a),
\end{equation}
where $L$ is the spatial extent of the smaller of the lattices at each
$\beta$.  The results of this analysis are provided in
Table~\ref{tab:finite_volume}; the mass of the negative-parity state
on the larger lattices is higher than on the smaller lattices at both
lattice spacings, by an amount of order 5\%.

\begin{figure}[tp]
\begin{center}
\epsfig{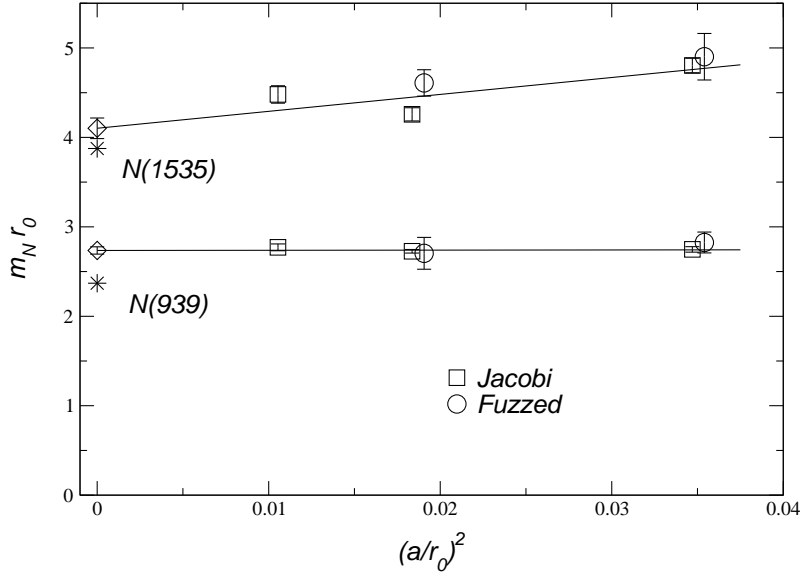}
\end{center}
\caption{The masses of the lowest-lying positive- and negative-parity
baryons in units of $r_0^{-1}$~\protect\cite{sommer94,wittig98}
against $a^2$ in units of $r_0^2$.  The lines are linear fits in
$a^2/r_0^2$ to the positive- and negative-parity baryon masses.  Also
shown are the physical values.}\label{fig:continuum_extrap}
\end{figure}

In order to look at the discretisation uncertainties in our data, we
show in Figure~\ref{fig:continuum_extrap} the masses in units of $r_0$
against the $a^2/r_0^2$, where $r_0 = 0.5~\mbox{fm}$ is the hadronic
scale~\cite{sommer94,wittig98}.  For the lightest positive-parity
state, we obtain entirely consistent results at the different lattice
spacings.  For the case of the negative-parity states, there is some
trend towards decreasing mass at finer lattice spacings.  In order to
quantify the discretisation uncertainties, we perform uncorrelated
$\chi^2$ fits in $a^2/r_0^2$:
\begin{equation}
(M_X r_0)(a) = M_X r_0 (a = 0) + c \left(\frac{a}{r_0}\right)^2,
\end{equation}
where $X$ is either $N^{1/2+}$ or $N^{1/2-}$, yielding
\begin{eqnarray}
M_{N^{1/2+}} r_0 & = & 2.74(4)\nonumber\\
M_{N^{1/2-}} r_0 & = & 4.1(1).\label{eq:final_result}
\end{eqnarray}
The fits and extrapolated masses are shown on the figure, together
with the experimental values of the masses in units of $r_0$.  We
estimate the difference between the values obtained in the continuum
limit and that obtained on our finest lattice, $\beta = 6.4$, as a
measure of the discretisation uncertainties in our calculation; this
is negligible for the positive-parity state, and 10\% for the negative
parity state.

Though there is a noticeable discrepancy between the lattice and
physical values, a direct experimental measurement of $r_0$ is
unavailable, and $r_0$ is a better scale for comparing data at
different lattice spacings than for comparing data with experiment.
Therefore we choose as our final result the mass ratio of the
negative- and positive-parity masses in the quenched approximation
\begin{equation}
M_{N^{1/2-}}/M_{N^{1/2+}} = 1.50(3),
\end{equation}
where the quoted error is purely statistical, and we estimate
systematic uncertainties of order 10\% due to finite volume effects,
and due to chiral extrapolation and discretisation uncertainties.
This ratio is to be compared with the physical ratio of 1.63.

\section{Discussion and Conclusions}\label{sec:concl}
The calculation exhibits a clear mass splitting between the positive-
and negative-parity states, in agreement with calculations using the
highly-improved and domain-wall fermion actions.  In view of the
dearth of studies of negative-parity baryon masses using the Wilson
fermion action, it is instructive to compare the mass splitting
obtained with the clover fermion action with that obtained from the
standard Wilson fermion action.  A discrepancy in the splitting
between the two actions would be indicative of lattice artifacts, that
would presumably be more severe in the case of the Wilson fermion action.  In
order to study this, Wilson quark propagators were computed on the
UKQCD $24^3\times 48$ lattices at $\beta = 6.2$ at a quark mass
corresponding to $m_{\pi}/m_{\rho} = 0.7$; the smearing and fitting
procedures were the same as those employed in the SW calculation.

\begin{figure}[tp]
\begin{center}
\epsfig{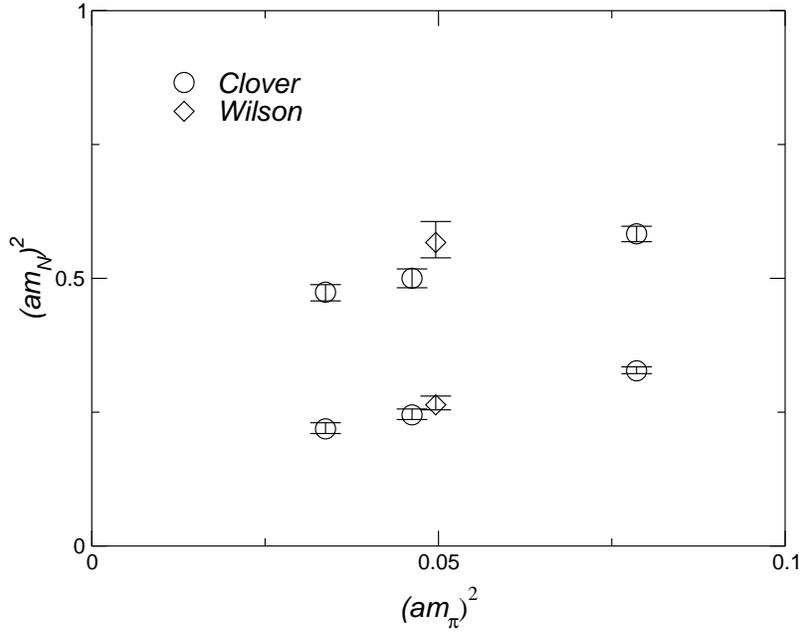}
\end{center}
\caption{The masses in lattice units of the lowest-lying positive- and
negative-parity nucleons using the SW-clover action at $\beta =
6.2$ on the $24^3\times 48$ lattices (circles).  Also shown is the
corresponding results obtained using the Wilson fermion action
(diamonds) on the same ensemble of configurations.}
\label{fig:wilson_vs_clover}
\end{figure}

The splitting between the masses of the positive- and negative-parity
states obtained with the Wilson fermion action, together with that
obtained with the SW fermion action is shown in
Figure~\ref{fig:wilson_vs_clover}.  The splittings obtained with the
two actions are entirely consistent, which we will now argue is
reasonable.  Under the $SU(6)$ spin-flavour symmetry, the low-lying
negative-parity baryons, up to around $2 \mbox{GeV}$, can be assigned
to a $l = 1$ $\underline{70}$-plet.  Similarly, the
low-lying positive-parity baryons can be assigned to an $l = 0$
$\underline{56}$-plet.  Thus the splitting in the parity doublets
is analogous to the $P-S$ splitting in the meson sector, which we know
is relatively faithfully reproduced using the Wilson fermion action.
Indeed an earlier lattice calculation demonstrating that the $P$-wave
baryons are accessible to lattice calculation is contained in
ref.~\cite{degrand92}.

Within the $l=1$ $\underline{70}$-plet, calculations of the masses of
these states both in the quark model\cite{isgur_karl}, and in large
$N_C$\cite{carlson99}, suggest that the spin-orbit contribution is
surprisingly small, whilst the hyperfine contribution is of normal
size, and there is an effective interaction carrying the quantum
numbers of pion exchange~\cite{goity97}.  It is within a multiplet
that we might expect the choice of action to play a r\^{o}le.

\begin{figure}[tp]
\begin{center}
\epsfig{width=250pt,file=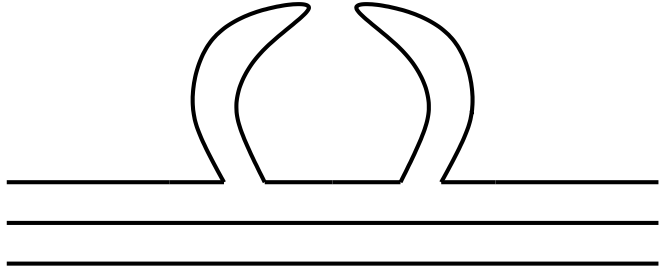}
\end{center}
\caption{Diagram contributing to the decay $N^{1/2-} \longrightarrow
N^{1/2+} \eta'$ in quenched QCD}
\label{fig:non_unitary}
\end{figure}

The continuation of the calculation to lighter quark masses will
require careful consideration.  A baryon can change parity through the
emission of a $\pi$ or $\eta'$; the latter process is accessible even
in the case of quenched QCD, where the $\pi$ and $\eta'$ are degenerate
in mass, as illustrated in Figure~\ref{fig:non_unitary}.\footnote{We
are grateful to Robert Edwards and Chris Michael for this
observation.} The non-unitary behaviour associated with such processes
has been observed in the scalar
correlator~\cite{thacker00,degrand01,duncan01}, and the presence of
such quenched oddities in the baryon sector discussed in
ref.~\cite{labrenz96}.  Whilst no evidence for non-unitary behaviour
is observed in this calculation, the lightest quark mass is indeed
close to the $N^* \rightarrow N \pi$ threshold.

In this letter it has been shown that both the SW-clover and Wilson
fermion actions are capable of resolving the splitting between the
positive- and negative-parity baryon masses in the quenched
approximation to QCD, and therefore that these are accessible to
relatively economical calculation.  We obtain a ratio for the masses
of the negative- and positive-parity states of $1.50(3)$, where the
error is purely statistical, compared with the experimental value of
$1.63$. A more extensive study of the spectrum including the case of
non-degenerate quark masses, and for the lowest-lying $I =
\frac{3}{2}$ states will appear in a longer paper~\cite{lhpc01}.

\section*{Acknowledgements}
This work was supported in part by DOE contract DE-AC05-84ER40150
under which the Southeastern Universities Research Association (SURA)
operates the Thomas Jefferson National Accelerator Facility, by the
European Community's Human Potential Program under Contract
HPRN-CT-2000-00145 (Hadrons/Lattice QCD), by EPSRC grant GR/K41663,
and PPARC grants GR/L29927 and GR/L56336.  CMM acknowledges PPARC
grant PPA/P/S/1998/00255. MG acknowledges financial support from the
DFG (Schwerpunkt ``Elektromagnetische Sonden''). Propagators were
computed using the T3D at Edinburgh, the T3E at ZIB (Berlin) and
NIC(J\"{u}lich), the APE100 at NIC (Zeuthen), and the \textit{Calico}
cluster at Jefferson Laboratory.

We are grateful for fruitful discussion with S.~Dytman, R.~Edwards,
J.~Goity, N.~Isgur, R.~Lebed, F.~Lee, C.~Michael, C.~Schat, A.~Thomas and S.~Wallace.

\begin{table}[tp]
\begin{center}
\begin{tabular}{cccccc}
$\beta$ & $\csw$ & $L^3\cdot T$ & $L\,[{\rm fm}]$ & $\kappa$ & Smearing \\
\hline
6.4 & 1.57  & $32^3\cdot 48$ & 1.6 & $0.1313, \, 0.1323,\, 0,1330, \, 0.1338,
        \, 0.1346, \, 0.1350$ & J\\[1.0ex]
6.2 & 1.61 & $24^3\cdot48$ & 1.6 & $0.1346,\,0.1351,\,0.1353$ & F \\
& &  $24^3\cdot 48$ & 1.6 & 
$0.1333,\, 0.1339, \, 0.1344,\, 0.1349, \, 0.1352$ &  J  \\
    &    & $32^3\cdot 64$ & 2.1 & $0.1352, \, 0.1353, \, 0.13555$ &
        J \\[1.0ex]
6.0 & 1.76 & $16^3\cdot48$ & 1.5 & $0.13344,\,0.13417,\,0.13455$ & F\\
& & $16^3\cdot 32$ & 1.5 & $0.1324,\, 0.1333, \, 0.1338, \, 0.1342$ &
        J\\
& & $24^3\cdot 32$ & 2.2 & $0.1342, \, 0.1346, \, 0.1348$ & J\\[1.0ex]
\end{tabular}
\end{center}
\caption{The parameters of the lattices used in the calculation.  The
labels $J$ and $F$ refer to use of Jacobi and ``fuzzed'' quark sources
respectively. Lattice sizes in physical units are quoted using $r_0$
to set the scale.\protect\cite{wittig98}}\label{tab:params}
\end{table}

\begin{table}[tp]
\begin{center}
\begin{tabular}{cccccc}
$\beta$ & 6.4 & \multicolumn{2}{c}{6.2} & \multicolumn{2}{c}{6.0}\\
 & Jacobi & Jacobi & Fuzzed & Jacobi & Fuzzed \\ \hline
$a M_{N^{1/2+}}$ & $0.285(4)$ & $0.369(3)$ & $0.37(2)$ & $0.512(6)$ &
 $0.53(2)$\\
$b_2$ & $2.38(2)$ & $2.40(2)$ & $2.5(3)$ & $2.42(3)$ & $2.3(2)$ \\ \hline
$a M_{N^{1/2-}}$ & $0.46(1)$ & $0.58(1)$ & $0.62(2)$ & $0.89(2)$ &
 $0.91(5)$\\
$b_2$ & $2.58(6)$ & $2.67(7)$ & $2.4(4)$ & $2.1(2)$ & $1.7(7)$ \\
\end{tabular}
\end{center}
\caption{The parameters of the fit of the lightest positive- and
negative-parity states to
eqn.~(\ref{eq:chiral_extrap}).}\label{tab:chiral_extrap}
\end{table}

\begin{table}
\begin{center}
\begin{tabular}{cccc}
$\beta$ & $a m_{\pi}$&  $a m_{N^{1/2-}}$ (small) &
$a m_{N^{1/2-}}$ (large)\\ \hline
6.0 & 0.3 & 0.97(1) & 1.01(1)\\
6.2 & 0.2 & 0.664(9) & 0.69(4)
\end{tabular}
\end{center}
\caption{A comparison of the masses of the lowest-lying negative-parity
states on the small-volume and large-volume lattices at $\beta = 6.0$
and $\beta = 6.2$, at a pseudoscalar mass $a m_{\pi} = 4.8/(L/a)$.}
\label{tab:finite_volume}
\end{table}

\vspace{0.5cm}

\vspace{0.5cm}
\vspace{0.5cm}
\vspace{0.5cm}
\vspace{0.5cm}

\end{document}